\documentclass[12pt,,letterpaper]{JHEP3}
\usepackage{graphics}
\usepackage{epsfig}
\usepackage{slashed}
\usepackage{amsmath}
\usepackage{subfigure}
\usepackage{cite}
\def\be{\begin{equation}}
\def\ee{\end{equation}}
\def\beq{\begin{equation}}
\def\eeq{\end{equation}}
\def\ba{\begin{array}}
\def\ea{\end{array}}
\def\zz{\mathcal{Z}}
\def\cc{\dot{\mathcal{C}}}
\newcommand{\bea}{\begin{eqnarray}}
\newcommand{\eea}{\end{eqnarray}}

\title{Complexity growth rate, grand potential and partition function}
\author{Wei Sun${}^{1}~$, Xian-Hui Ge${}^{1,2}~$,   \\
${}^{1}$Department of Physics, College of Sciences, Shanghai University, Shanghai 200444, People's Republic of China\\
${}^{2}$Shanghai Key Laboratory of
High Temperature Superconductors, Shanghai University, Shanghai 200444, People's Republic of China\\
{\sf{gexh@shu.edu.cn}}
}

\abstract{We examine the complexity/volume conjecture and further investigate the possible connections between complexity and partition function.  The complexity/volume 2.0 states that the complexity growth rate $\mathcal{\dot{C}}\sim PV$. In the standard statistics, there is a fundamental relation among $PV$, the grand potential $\Omega$ and the partition function $\mathcal{Z}$. By using this relation, we are  able to construct an ansatz between complexity and partition function. The complexity/partition function relation is then utilized to study the complexity of the thermofield double state of extended SYK models for various conditions. The relation between complexity growth rate and black hole phase transition is also discussed.
 }

\begin{document}
 \section{Introduction}
     The Anti-de Sitter/Conformal field theory (AdS/CFT)\cite{Maldacena:1997re,Gubser:1998bc,Witten:1998qj,Aharony:1999ti} duality states that the gravity theory of the AdS space-time can be described by the conformal field theory on the boundary. This not only offers us a new way to calculate the physical quantities on the field theory side, but also provides us with a new ideas for understanding the nature of space-time. In particular, according to the holographic entanglement entropy \cite{Ryu:2006bv}, there is a basic connection between quantum information theory and gravitational physics. However, in view of the thermo-field double state (TFD state) of the eternal black hole, it has been proved that entanglement entropy cannot provide us all the information in the evolution of the AdS wormhole \cite{Maldacena:2001kr,Hartman:2013qma}. The Einstein-Rosen Bridge (ERB) usually connects the two sides of the Penrose diagram of the eternal AdS black hole, and classically it will grow forever. On the other hand, the dual TFD state on the boundary reaches its thermal equilibrium very quickly. So how to describe the continuing growth of ERB for a long time when the quantum states on both two sides stop evolving in the dual theory? To solve this problem, Susskind and his collaborators \cite{Susskind:2013aaa,Susskind:2014rva,Susskind:2016tae,Stanford:2014jda} proposed a new concept called quantum computational complexity of black hole which can describe the quantum evolution of the boundary state after reaching thermal equilibrium. This concept can help us with some useful tools on studying problems of quantum complexity \cite{Brown:2015bva}.

     Note that Maldacena and Susskind have established a connection between Einstein-Podolsky-Rosen (EPR) in quantum mechanics and the Einstein-Rosen bridge (ERB) in gravity (so-called ER = EPR) \cite{Maldacena:2013xja,Susskind:2014yaa}. Based on the above conjecture, Alice on one side of the ERB can establish communication with Bob on the other side, but how difficult is it ? Quantum computational complexity can be understood as a candidate quantity to characterize how difficult it is in the calculation. In quantum circuits \cite{Hayden:2007cs},
     complexity is usually defined as the minimal number of gates used for processing the unitary operation\cite{Susskind:2014rva}. Susskind related computational complexity to the distance from the layered stretched horizon in \cite{Susskind:2013aaa}, and further proposed a conjecture that the length of the ERB is proportional to complexity of quantum state of the dual CFT.
     Inspired by the work of Hartman and Maldacena \cite{Hartman:2013qma}, Susskind and Stanford proposed a new version that the complexity is dual to the volume of the maximal spatial slice crossing the ERB called Complexity-Volume (CV) duality\cite{Stanford:2014jda}
     \begin{equation}
     \mathcal{C}\sim\frac{V}{G l_{AdS}},
     \end{equation}
     where $l_{AdS}$ is the length scale that has to be chosen appropriately for the configuration. While this proposal captures the linear growth at late time, there is also a minor problem that length scales must be introduced manually. In a recent work \cite{Brown:2015bva}(see \cite{Brown:2015lvg} for details), Susskind further proposed an alternative conjecture that the quantum complexity of a holographic state is dual to the action of certain Wheeler-DeWitt (WDW) patch in the AdS bulk so-called Complexity-Action (CA)
     \begin{equation}
     \mathcal{C}=\frac{A}{\pi\hbar},
     \end{equation}
     where $A$ is the action of the Wheeler-DeWitt patch.  This proposal solves the length scale problem of CV-duality and has the practical advantage that the WDW patch is easier to work with than the maximal volume. It was noted that both the CV duality and the CA duality share the same properties as follows \cite{Brown:2015lvg}:

     \textit{The rate of complexity growth is bounded by the product of entropy and temperature $\frac{d \mathcal{C}}{dt}\sim T S$}.

      There have been many studies on CV duality and CA duality, such as the divergence structure \cite{Carmi:2016wjl,Kim:2017lrw}, complexity growth rate\cite{Lehner:2016vdi,Miao:2017quj,Carmi:2017jqz,An:2018xhv,Cai:2017sjv,Jiang:2019qea,Ghaffarnejad:2018prc,roy18}, and the generalization beyond Einstein gravity \cite{Cai:2016xho,Jiang:2018pfk,Cano:2018aqi,An:2018dbz,Jiang:2019fpz}(see also \cite{Fan:2019mbp,Couch:2016exn,Fan:2018wnv}), For example, some of us have tried to relate complexity with the phenomena of the accelerating expansion of our universe \cite{Ge:2017rak}. There is also a study on the complexity of disk-shape subregion in various (2+1)-dimensional gapped systems with gravity dual \cite{duwu2018} . Given the fact that many literatures \cite{Kastor:2009wy,Kubiznak:2014zwa,Dolan:2012jh,Frassino:2015oca,Kubiznak:2016qmn,Johnson:2014yja} have taken the cosmological constant as pressure, an improved version of the CV conjecture was proposed in \cite{Couch:2016exn} called ``complexity=volume 2.0" (CV 2.0)
     \begin{equation}
     \mathcal{C}\sim \frac{1}{\hbar}P (Spacetime \  Volume)   .
     \end{equation}
     In the late time regime, it was proposed that
     \begin{equation}
     \dot{\mathcal{C}}\sim\frac{PV}{\hbar}\label{c10} .
     \end{equation}
     The above equation relates complexity to pressure and thermodynamic volume. The authors in \cite{Couch:2016exn} claimed that the CA duality would violate the Lloyd bound in some cases of charged black hole, while CV 2.0 would not, which shows the rationality of this conjecture. Subsequently, it was proposed CA2.0 \cite{Fan:2018wnv}
     \begin{equation}\label{scope}
     \mathcal{C}=\frac{A_{\Lambda}}{\pi\hbar} ,
     \end{equation}
     where $A_{\Lambda}$ is a part of the non-derivative action evaluated on the WDW patch. The application scope of equation (\ref{scope}) is not limited to the late time of the black hole evolution. Especially $A_{\Lambda}$ reduces to PV in the stationary limit. That will return to CV2.0. In a very recent paper, a new CV duality was proposed $\dot{\mathcal{C}}=2P\Delta V$ in \cite{Liu:2019mxz}. The rationality of the various versions of complexity and the existence of more reasonable conjectures deserve further study.

    From the standard thermodynamics, we know the relation between $PV$ and the grand  potential $\Omega$
    \begin{equation}\label{pvm}
     pV=-\Omega.
     \end{equation}
    This relation together with (\ref{c10}) stimulates us to think about building some deep connections among the complexity growth rate, thermodynamical quantities and statistical physics.  Moreover, the comparison between ordinary thermodynamics and black hole thermodynamics is shown in table \ref{tableone}. When the system is in its thermodynamic equilibrium, $\Omega$ is at its minimum. The thermodynamical stability requires
    $d \Omega \leq 0$
     \footnote{This in turn implies $d \dot{C}\geq 0$. The variation of the complex growth rate then has a property similar to entropy. Actually,  the second law of complexity states \cite{Brown2018The}:
    \textit{If the computational complexity is less than maximum, then with overwhelming likelihood it will increase, both into the future and into the past. }}.
    \begin{table}\label{tableone}
      \centering
      \begin{tabular}{|c|c|c|c|}
     	\hline
     	&Thermodynamics&Neutral Black Hole&Charged Black Hole\\
     	\hline
     	First Law&dH=TdS+VdP&dM=TdS+VdP&dM=TdS+VdP+$\Phi$dQ\\
     	\hline
     	Enthalpy&H=U+PV&H=M=U+PV&H=M=U+PV\\
     	\hline
     	Free energy&F=U-TS&F=U-TS&F=U-TS\\
     	\hline
     	Gibbs free energy&G=H-TS&G=M-TS&G=M-TS\\
     	\hline

      \end{tabular}
      \caption{Ordinary thermodynamic relationship and black hole thermodynamic relationship}
     \end{table}
  For canonical ensembles, the grand potential reduces to the free energy $F$. In ordinary thermodynamics, the principle of  maximum work states that:

   \textit{For all thermodynamic processes between the same initial and final state, the delivery of work is a maximum for a reversible process, obeying $dW\leq -dF$. }

   Quantum complexity is a kind of computational resource. The quantum computational process can be regarded as a thermodynamic process in which work should be delivered. Less complexity indicates less time or work is required. Then equations (\ref{c10}) and (\ref{pvm}) together with the principle of maximum work indicates that complexity may related to the free energy of the system.  Moreover, this could lead to $d \mathcal{\dot{C}} \leq -dF $.

     On the other hand, by studying the thermodynamics of black holes, we can understand the thermodynamic behavior of strongly coupled field theory systems at finite temperatures, while traditional quantum field theory is hard to work. Black hole phase transition is an important part of black holes. The famous Hawking-Page phase transition\cite{Haking:1983} corresponds to the confinement and deconfinement phase transition in field theory. As complexity can be related to the grand potential, it would be interesting to study the black hole phase transition by using complexity as a probe. We can also test the rationality of this version of the conjecture by calculating the evolutionary behavior of complexity over time in the phase transition process. On the other hand, we can also reconstruct the space-time in the bulk from the nature of the boundary complexity.

     In this paper, our main purpose is to examine the universality of CV2.0 by establishing a connection between this conjecture,  the grand  potential and the grand partition function. The reasons for connecting CV 2.0 with the partition function are largely due to the fact that complexity of the TFD state of various extended Sachdev-Ye-Kitaev (SYK) model calls for further investigation. Being one of the simplest strongly interacting system with a gravity dual, the SYK model has many appealing features including thermodynamical and transporting properties. These properties suggest that the SYK models are connected holographically to black holes with nearly $AdS_2$ horizons. The operator complexity of the SYK model has been studied in \cite{rqyang19} in which it was concluded that the complexity grows linearly. We are going to investigate the complexity of the corresponding TFD state of various deformations of the SYK model by exploring the complexity/partition function relation. This may provide additional evidence supporting the SYK model/gravity duality.

     The structure of this paper is organized as follows. In section 2, we will establish the connection between complexity, the grand potential and the corresponding partition function.
      We take various deformations of the SYK model as concrete examples and calculate the corresponding complexity growth rate of the corresponding TFD state. Then, in section 3, we extend our discussions to Schwarzschild-AdS and Reissner-Nordstrom AdS black holes. In section 4, we relate complexity growth rate to black hole phase transitions since the grand potential $\Omega$ can describe phase transitions.  In section 5, we investigate whether our proposal violates the Lloyd bound. The conclusions and discussions are provided in the last section.

 \section{Complexity and partition function}
    In the standard statistical physics, the grand thermodynamic potential is closely related to the grand partition function $\mathcal{Z}$ via \cite{Pathria1996Statistical}
    \begin{equation}\label{partition}
    \Omega=-k T \ln \mathcal{Z}.
    \end{equation}
    From the ansatz $\dot{C}\sim pV/\hbar\sim -\Omega/\hbar$, we have
     \begin{equation}
    \dot{C}= \frac{k T}{\hbar} \ln \mathcal{Z}.
    \end{equation}
    We refer this ansatz as  ``complexity growth rate/partition function relation".  An alternative approach to obtain the relation between $\mathcal{Z}$ and $pV$ is as follows. The partition function is given by
     \begin{equation}
    \mathcal{Z}= e^{\sum_{r,s}(-\alpha N_r-\beta E_s)},
    \end{equation}
    where $\alpha=\frac{\mu}{kT}$, $\beta=\frac{1}{kT}$, $N_r$, $E_s$ and $\mu$ denote the particle number, energy and chemical potential of a system, respectively. Note that
    \begin{equation}
    \label{twofour}
    d\ln\mathcal{Z}= -\bar{N}d\alpha-\bar{E}d\beta-\frac{\beta}{\mathcal{N}}\sum_{r,s}\langle n_{r,s}\rangle dE_s,
    \end{equation}
    where the averaged particle number and energy are given by
     \bea
    \bar{N}&\equiv&-\frac{\partial}{\partial \alpha}\ln \mathcal{Z}= \frac{\sum_{r,s}N_r e^{(-\alpha N_r-\beta E_s)}}{\sum_{r,s}e^{(-\alpha N_r-\beta E_s)}},\\
    \bar{E}&\equiv& -\beta\frac{\partial}{\partial \beta}\ln \mathcal{Z}.
    \eea
   Equation (\ref{twofour}) can be recast as
    \be
    d(\ln \mathcal{Z}+\alpha \bar{N}+\beta \bar{E})=\beta\bigg(\frac{\alpha}{\beta}d\bar{N}+d\bar{E}-\frac{1}{\mathcal{N}}\sum_{r,s}\langle n_{r,s}\rangle dE_s\bigg).
    \ee
    In comparison with the first law of thermodynamics
    \be
    \delta Q=d\bar{E}+\delta W-\mu d\bar{N},
    \ee
    we arrive at
    \be
    \beta \delta Q=d(\ln \mathcal{Z}+\alpha \bar{N}+\beta \bar{E})=\frac{S}{k}.
    \ee
    Therefore, we obtain
    \be
    \ln \mathcal{Z}=\frac{S}{k}-\alpha \bar{N}-\beta \bar{E}.
    \ee
    By further using the relation $G=\bar{E}-TS+pV$, we finally obtain
    \begin{equation}
     \ln\mathcal{Z}=\frac{PV}{kT}\label{commutative2} .
     \end{equation}
     This is a fundamental relation between thermodynamics and statistic physics.
      Connecting quantum complexity to partition functions through (\ref{c10}) and (\ref{commutative2}) goes beyond the original ``CV 2.0" conjecture. One may call it ``Complexity/Grand potential/Partition Function" relation or simply ``CV 3.0"
     \begin{equation}
     \ln\mathcal{Z}=\frac{PV}{kT}\sim\frac{\dot{\mathcal{C}}{\hbar}}{kT}\label{commutative3} .
     \end{equation}
    This formula is able to relate complexity closely to the microscopic physics of the SYK model and black holes. We will evaluate this formula for various deformations of the SYK model and black holes. Hereafter, we take $\hbar=k=1$.

    \subsection{The complexity/partition function relation in the SYK model}
    The SYK model as a quantum many-body model has many beautiful structures and properties similar to a black hole. It is solvable in the large $N$ limit and the low-energy limit of the SYK model leads to a nonconformal contribution to four-point functions captured by a Schwarzian derivative.  In this section, we examine the CP conjecture for the SYK model. The original SYK model has been studied in \cite{kitaev15,Maldacena:2016hyu,davison16}. We are going to examine the complexity/partition function relation by utilizing the partition function given in \cite{kitaev15,Maldacena:2016hyu}.

    The SYK model is a quantum-mechanical model with $N$ Majorana fermions with random interactions involving $q$ of these fermions at a time, where $q$ is an even number. The Hamiltonian is \cite{Maldacena:2016hyu} \begin{equation}
    H=(i)^{q/2}\sum_{1\leq i_1<i_2...i_{q}\leq N}j_{i_1 i_2...i_q}\chi_{i_1}\chi_{i_2}\cdots \chi_{i_q}.
    \end{equation}
    Each coefficient is a real variable drawn from a random Gaussian distribution satisfies $\langle j^2_{i_1\ldots i_q}\rangle=J^2 (q-1)!/N^{q-1}$.
    After writing the original partition function of the theory as a functional integral with a collective action based on the Luttinger-Ward analysis \cite{georges01}, one can obtain the free energy and the entropy. The general expression of the free energy, in a low temperature expansion, has the form \cite{kitaev15}
    \begin{equation}
    \log \mathcal{Z}=-\beta E_{0}+S_{0}+\frac{c}{2\beta}+\cdots,
    \end{equation}
    where the ground state energy, entropy and specific heat are all proportional to $N$.
    The zero temperature entropy is given for general $q$, that is
    $     \frac{S_{0}}{N}\sim \frac{1}{2}\log 2-\frac{\pi^{2}}{4q^{2}}+\cdots$.
 From the relation $\dot{\mathcal{C}}\sim T \ln \zz$, the complexity growth rate is then given by
    \begin{equation} \label{rate}
    \dot{\mathcal{C}} \sim -E_0+\frac{S_0}{\beta}+\frac{c}{2\beta^2}+...
    \end{equation}
    The first two terms in (\ref{rate}) are consistent with the complexity of charged black holes obtained in \cite{Susskind:2014rva,Brown:2015lvg}, while the third term can be considered as a higher order correction. Equation
     (\ref{rate}) reflects that $TS_0$ is competing with the ground energy $E_0$. This also agrees with the behavior one would expect based on a quantum circuit model of complexity
      \cite{Susskind:2014rva,Hayden:2007cs}: The rate of quantum computation measured in gates per unit time is proportional to the product $T S$; The entropy appears because it represents
      the width of the circuit and the temperature is an obvious choice for the local rate at which a particular qubit interacts.


    \subsection{The complex SYK model}

    We can extend our discussion to include the case of the complex SYK fermions. The zero-dimensional SYK model with complex fermions $f_i$ label by $i=1,...N$. The Hamiltonian is \cite{davison16}
     \be
    H_0=\sum_{1\leq i_1<i_2...i_{q/2}\leq N}J_{i_1 i_2...i_q}f^{\dagger}_{i_1}f^{\dagger}_{i_2}\cdot\cdot\cdot f^{\dagger}_{i_{q/2}}f_{i_{q/2+1}}\cdot\cdot\cdot f_{i_{q-1}}f_{i_{q}}.
    \ee
    The grand potential is given by \cite{davison16}
    \be
    \Omega=...-J^2\int^{1/T}_0 d\tau [G(\tau)]^{q/2}[G(1/T-\tau)]^{q/2}.
    \ee
    The thermodynamics of the zero-dimensional complex SYK model was discussed in \cite{davison16}.
    The complete grand potential, including the contribution of the ground state energy is given by,
    \be
    \Omega=E_0-\mu_0 \mathcal{Q}-T \mathcal{G}+...,
    \ee
    where $\mathcal{Q}$ is the charge density, $\mathcal{Q}$ is a parameter related to the entropy. That is to say
    \bea
    \mathcal{Q}&=&-\frac{1}{2\pi}\frac{d \mathcal{G}}{d\mathcal{E}},\\
    \mathcal{S}&=&\mathcal{G}+2\pi \mathcal{E} \mathcal{Q},
    \eea
    where $\mathcal{S}$ is the entropy and $\mathcal{E}$ is a parameter controlling the particle-hole symmetry.
    Subtracting the ground state energy, we simply obtain
    \be
    \dot{\mathcal{C}}\sim T \mathcal{G}.
    \ee
    In the $\mathcal{E}\rightarrow 0$ limit, it becomes $\dot{\mathcal{C}}\sim T \mathcal{S}$. This result also agrees with \cite{Brown:2015lvg}.

    \subsection{Complexity growth rate and higher dimensional SYK model}
    Higher dimensional extensions of the original SYK models were investigated widely because such models can give interesting quantum critical properties, such as linear-in-T resistivity \cite{song2017,Chowdhury:2018sho}, many-body localization to metal phase transition \cite{yao2017,whcai2018} and so on. Recently, Patel et al. \cite{Patel2018} and Chowdhury et al. \cite{Chowdhury:2018sho} constructed a (2+1)dimensional strongly correlated solvable model, consisting of coupled SYK islands, which yields linear-in-T and linear-in-B behaviors.
    We are going to  examine the complexity growth rate of this model.
     The (2+1)-dimensional SYK model given in \cite{Chowdhury:2018sho} with the Hamiltonian
    \begin{equation}
    H=H_{c}+H_{f}+H_{cf},
    \end{equation}
    with
    \begin{equation}
    H_{c}=\sum_{\boldsymbol r,\boldsymbol r^{'}} \sum_{l}( -t^{c}_{\boldsymbol r,\boldsymbol r^{'}}-\mu_{c}\delta_{\boldsymbol r,\boldsymbol r^{'}} ) c^{\dagger}_{\boldsymbol r  l}c_{\boldsymbol r^{'}  l}+\frac{1}{(2N)^{3/2}}\sum_{\boldsymbol r}\sum_{ijkl}u^{c}_{ijkl}c^{\dagger}_{\boldsymbol r  i}c^{\dagger}_{\boldsymbol r  j}c_{\boldsymbol r  k}c_{\boldsymbol r  l},
    \end{equation}
    \begin{equation}
    H_{f}=\sum_{\boldsymbol r,\boldsymbol r^{'}} \sum_{l}( -t^{f}_{\boldsymbol r,\boldsymbol r^{'}}-\mu_{f}\delta_{\boldsymbol r,\boldsymbol r^{'}} ) f^{\dagger}_{\boldsymbol r  l}f_{\boldsymbol r^{'}  l}+\frac{1}{(2N)^{3/2}}\sum_{\boldsymbol r}\sum_{ijkl}u^{f}_{ijkl}f^{\dagger}_{\boldsymbol r  i}f^{\dagger}_{\boldsymbol r  j}f_{\boldsymbol r  k}f_{\boldsymbol r  l}.
    \end{equation}
    The inter-band interaction $H_{cf}$ is chosen to be
    \begin{equation}
    H_{cf}=\frac{1}{N^{3/2}}\sum_{\boldsymbol r}\sum_{ijkl}V_{ijkl}c^{\dagger}_{\boldsymbol r  i}f^{\dagger}_{\boldsymbol r  j}c_{\boldsymbol r  k}f_{\boldsymbol r  l},
   \end{equation}
    where the coefficients  $V_{ijkl}$, are chosen to be identical at every site with $\overline{u^{f}_{ijkl}}=\overline{V_{ijkl}}=0 $, and the distribution of the couplings satisfy $\overline{(u^{f}_{ijkl})^{2}}=u^{2}_{f}$, $\overline{(u^{f}_{ijkl})^{2}}=0$ and $\overline{(v_{ijkl})^{2}}=u^{2}_{cf}$.  This model can be
regarded as two independent subsystems: the conducting
$c$ fermions with a hopping $t^{c}_{\boldsymbol r,\boldsymbol r^{'}}$, and the local and immobile $f$ fermions with SYK interaction at each site.
    As to the thermodynamics properties of the intermediate non-fermi liquid regime, one can evaluate the entropy density through $S=-\frac{\partial F }{\partial T}$. This gives three contributions to the entropy density $\mathcal{S}=\frac{S}{2NV}$,
    \begin{equation}
    \mathcal{S}(T)=\mathcal{S}_{c}(T)+\mathcal{S}_{f}(T)+\mathcal{S}_{int}(T),
    \end{equation}
    with
    \begin{equation}
    \mathcal{S}_{f}(T)=\mathcal{S}_{0,q}+\gamma_{q}T,
    \end{equation}
    \begin{equation}
    \mathcal{S}_{c}(T)\sim T^{1/z}\sim T^{4\Delta (q)},
    \end{equation}
    \begin{equation}
    \mathcal{S}_{int}(T)\sim T^{1+4\Delta (q)}.
    \end{equation}
    where $\Delta=\frac{1}{q}$ and $\gamma_q$ is a constant.
    Here, $\mathcal{S}_{f}(T)$ is the entropy of a single $SYK_{g}$ model, $S_{c}(T)$ comes from c-fermions, and $\mathcal{S}_{int}(T)$ originates from the inter-species interaction term $H_{int}$.
    The complexity growth rate is then
    \begin{equation}
    \dot{\mathcal{C}_{f}}\sim \displaystyle{ \int \mathcal{S}_{f} dT }=\displaystyle{ \int (\mathcal{S}_{0,q}+\gamma_{q}T) dT }=\mathcal{S}_{0,q}T+\frac{1}{2}\gamma_{q}T^{2}+\cdots,
    \end{equation}
    \begin{equation}
    \dot{\mathcal{C}_{c}}\sim \displaystyle{ \int \mathcal{S}_{c} dT }=\displaystyle{ \int  T^{4\Delta (q)} dT }=\frac{1}{4\Delta+1}T^{4\Delta+1}+\cdots,
    \end{equation}
    \begin{equation}
    \dot{\mathcal{C}}_{int}\sim \displaystyle{ \int \mathcal{S}_{int} dT }=\displaystyle{ \int  T^{1+4\Delta (q)} dT }=\frac{1}{4\Delta+2}T^{4\Delta+2}+\cdots.
    \end{equation}
The complexity growth rate of $f$ fermions obeys the relation $\dot{\mathcal{C}} \sim T \mathcal{S}_{0,q}$. However, the complexity growth rate for $c$-fermions and the interaction term do not obey on the entropy, which is because the $c$-fermions do not obey the SYK interaction. Therefore the subsystem for c-fermions does not yield a gravity dual. This result in turn further indicates that the SYK model indeed has its own gravity dual.

\subsection{Complexity growth rate of thermofield double state of SYK ``wormholes" }

 A pair of SYK islands
of Majorana fermions with identical two-body interactions, coupled by one-body hopping, have been used to
describe eternal traversable wormholes in a dual gravity theory \cite{maldacena201810}. The configuration contains negative null energy generated by quantum fields under the influence of an external coupling\cite{maldacena201810}. The dynamics of the two coupled SYK systems looks like that of a traversable wormhole.

The Hamiltonian takes the form \cite{maldacena201810}
\be
H_{\rm total}=H_{\rm L, SYK}+H_{\rm R, SYK}+H_{\rm int},~~~H_{\rm int}=i\mu\sum_{j}\chi^j_L\chi^j_R.
\ee
The system will develop an approximate conformal symmetry at energy scales less than $\mathcal{J}$. The effects of the coupling $\mu$ are as a perturbation to the approximately conformal system. The thermofield double state is a pure state of the combined, while left and right systems have a large value of the left-right correlators. At small coupling $\mu$, the ground state is very close to the thermofield double state $\rm |TFD\rangle$ of the decoupled systems.

At higher temperature, the partition function of the coupled system is then given by \cite{maldacena201810}
\be
\log \mathcal{Z}=2S_0+\frac{(2\pi)^2}{{\beta}}+\eta^2 \beta^{2-4\Delta}\int^1_0 dx \frac{\pi}{\sin x\pi}+...
\ee
where $S_0$ is the ground state entropy of each SYK model and $\eta$ is a parameter. To leading order, the complexity growth rate is given by
\be
\mathcal{\dot{C}}\sim  2S_0 T+\mathcal{O}(T^2).
\ee
This again agrees with our original proposal. However there is a factor difference from the result obtained on the gravity side, the complexity growth rate of JT gravity \cite{JTcomplexity}
\be
\mathcal{\dot{C}}\sim 4 T S_0+\mathcal{O}(T^2).
\ee
But our result agrees with \cite{cai2019}.

 \section{ The complexity/partition function relation for AdS black holes}

    The relation $\ln \mathcal{Z}\sim pV/T \sim \dot{\mathcal{C}}/T$ closely relates the microscopic physics ( i.e. the grand partition function) with the complexity growth rate. The origin of the microscopic states of Schwarzschild black hole still remains elusive. Within a semiclassical regime we can think of the partition function of the bulk theory as a path integral over metrics.  Given the Euclidean saddle points of the bulk theory, the partition function is
    \be
    \mathcal{Z}=e^{-I_{E}[g_{*}]},
    \ee
    where $I_{E}[g_{*}]$ is the Euclidean action at the saddle point. In the following, we use the notations given in \cite{hartnoll0903}. The bulk action for Schwarzschild-AdS black hole with the Gibbons-Hawking boundary term is given by
    \be
    I_{E}=-\frac{1}{2\kappa^2}\int_{M} d^{d+1}x\sqrt{g}\bigg(R+\frac{d(d-1)}{L^2}\bigg)+\frac{1}{2\kappa^2}\int_{\partial M}d^{d}x\sqrt{\gamma}\bigg(-2K+\frac{2(d-1)}{L}\bigg),
    \ee
    where $\gamma$ is the induced metric on the boundary and $K$ is the trace of the extrinsic curvature.

    One  saddle is obtained by analytic continuation of the Schwarzschild-AdS metric,
setting $\tau = it$. That is
\bea
ds^2_{*}&=&\frac{L^2}{r^2}\bigg[-f(r)dt^2+\frac{dr^2}{f(r)}+dx^i dx^i\bigg],\\
f(r)&=&1-\bigg(\frac{r}{r_{+}}\bigg)^d.
\eea
The Hawking temperature of the black hole is
\be
T=\frac{d  r_{+}}{4\pi L^2}.
\ee
The corresponding entropy is then given by
\be
S=\frac{(4\pi)^d L^{d-1}}{2\kappa^2 d^{d-1}}V_{d-1}T^{d-1}.
\ee
After some calculation, one can evaluate the action of the Euclidean Schwarzschild-AdS black hole
\be
I_{E}=-\frac{(4\pi)^d L^{d-1}}{2\kappa^2 d^d}V_{d-1}T^{d-1}.
\ee
The complexity growth rate then is given by
\be
\dot{\mathcal{C}}\sim -TI_{E}d=\frac{(4\pi)^d L^{d-1}}{2\kappa^2 d^{d-1}}V_{d-1}T^d
\ee
We can also obtain the free energy from the Euclidean action
\be
F=-T \ln \zz=-\frac{(4\pi)^d L^{d-1}}{2\kappa^2 d^d}V_{d-1}T^d.
\ee
We conclude that $F=-\dot{\mathcal{C}}=TS/d$.

For RN-AdS black holes, the corresponding bulk action is the Einstein-Maxwell theory
\be
I_{E}=\int d^{d+1}x\sqrt{g}\bigg[\frac{1}{2\kappa^2}\bigg(R+\frac{d(d-1)}{L^2}\bigg)-\frac{1}{4 g^2}F^2\bigg],
\ee
where $F=dA$ is the electromagnetic field strength. Working in the grand canonical ensemble with $\mu$ fixed and using the notation $\Omega=-T\ln \mathcal{Z}$, where $\zz$ is the partition function defined by the gravitational integral.
The metric of the RN-AdS black hole is given by
\bea
ds^2&=&\frac{L^2}{r^2}\bigg[-f(r)dt^2+\frac{dr^2}{f(r)}+dx^i dx^i\bigg],\\
f(r)&=&1-\bigg[1+\frac{r^2_{+}\mu^2}{\gamma^2}(\frac{r}{r_{+}})^d+\frac{r^2_{+}\mu^2}{\gamma^2}(\frac{r}{r_{+}})^{2(d-1)}\bigg].
\eea
The corresponding Hawking temperature is given by
\be
T=\frac{1}{4\pi r_{+}}\bigg(d-\frac{(d-2)r^2_{+}\mu^2}{\gamma^2}\bigg), ~~\gamma^2=\frac{(d-1)g^2L^2}{(d-2)\kappa^2}.
\ee
The grand potential is obtained as
\be
\Omega=-\frac{L^{d-1}}{2\kappa^2 r^d_{+}}\bigg(1+\frac{r^2_{+}\mu^2}{\gamma^2}\bigg)V_{d-1}.
\ee
From the complexity/grand potential relation, the complexity growth rate can then be obtained as
\be
\dot{\mathcal{C}}=-\Omega.
\ee
     The most remarkable feature of the holographic principle is the Bekenstein Hawking area law for black hole entropy
     \be
     S = \frac{A}{4}.
     \ee
    In thermodynamics, the well known thermodynamic relation is
     \begin{equation}
     S=-(\frac{\partial \Omega}{\partial T})_{\mu}.
     \end{equation}
     This ansatz strongly indicates that $ S=\frac{\partial \dot{\mathcal{C}}}{\partial T}$, so one may evaluate the complexity growth rate via $\dot{\mathcal{C}}=\int S dT$.

    Actually, the relation $\ln \mathcal{Z}\sim pV/T \sim \dot{\mathcal{C}}/T$ has its deep connections with the CA conjecture.
     In the Euclidean coordinates, the action $I_{E}$ is related to the partition function via
     \begin{equation}
     I_{E}=-\ln \mathcal{Z},\label{12}
     \end{equation}
     Therefore, we have
     \be
    \cc=-T I_{E}.
     \ee
     That is to say, complexity is closely related to the action in the Euclidean spacetime. The difference between the original CA conjecture and the formula obtained in (\ref{12}) is that there is a minus sign. This means that the complexity growth rate can be positive or negative, which actually relates to the stability of black holes. For thermodynamically stable black holes, the minimum and negative free energy refers that the complexity growth rate is positive. For thermodynamically unstable black holes, the corresponding positive free energy indicates that the complexity growth rate is negative signalizing phase transitions would happen.

 \section{Complexity growth rate and black hole phase transition}
  \subsection{Schwarzschild-AdS Black Hole}
   In this section we will focus on the relationship between complexity and AdS black hole phase transition. We focus on the 4-dimensional Schwarzschild-AdS black holes. For such black holes, $P=\frac{-{\Lambda}}{8\pi}$, $\Lambda=-\frac{3}{l^{2}}$, $V=\frac {4\pi r^{3}_{+}}{3}$, the standard Schwarzschild-AdS Black Hole metric is written as
     \begin{equation}
     ds^{2}=-\bigg(1-\frac{2M}{r}+\frac{r^{2}}{l^{2}}\bigg)dt^{2}+\bigg(1-\frac{2M}{r}+\frac{r^{2}}{l^{2}}\bigg)^{-1}dr^{2}+r^{2}d\Omega^{2} ,
     \end{equation}
     \begin{figure}[htbp]
    	\centering
    	\includegraphics[width=0.8\linewidth,height=0.4\textheight]{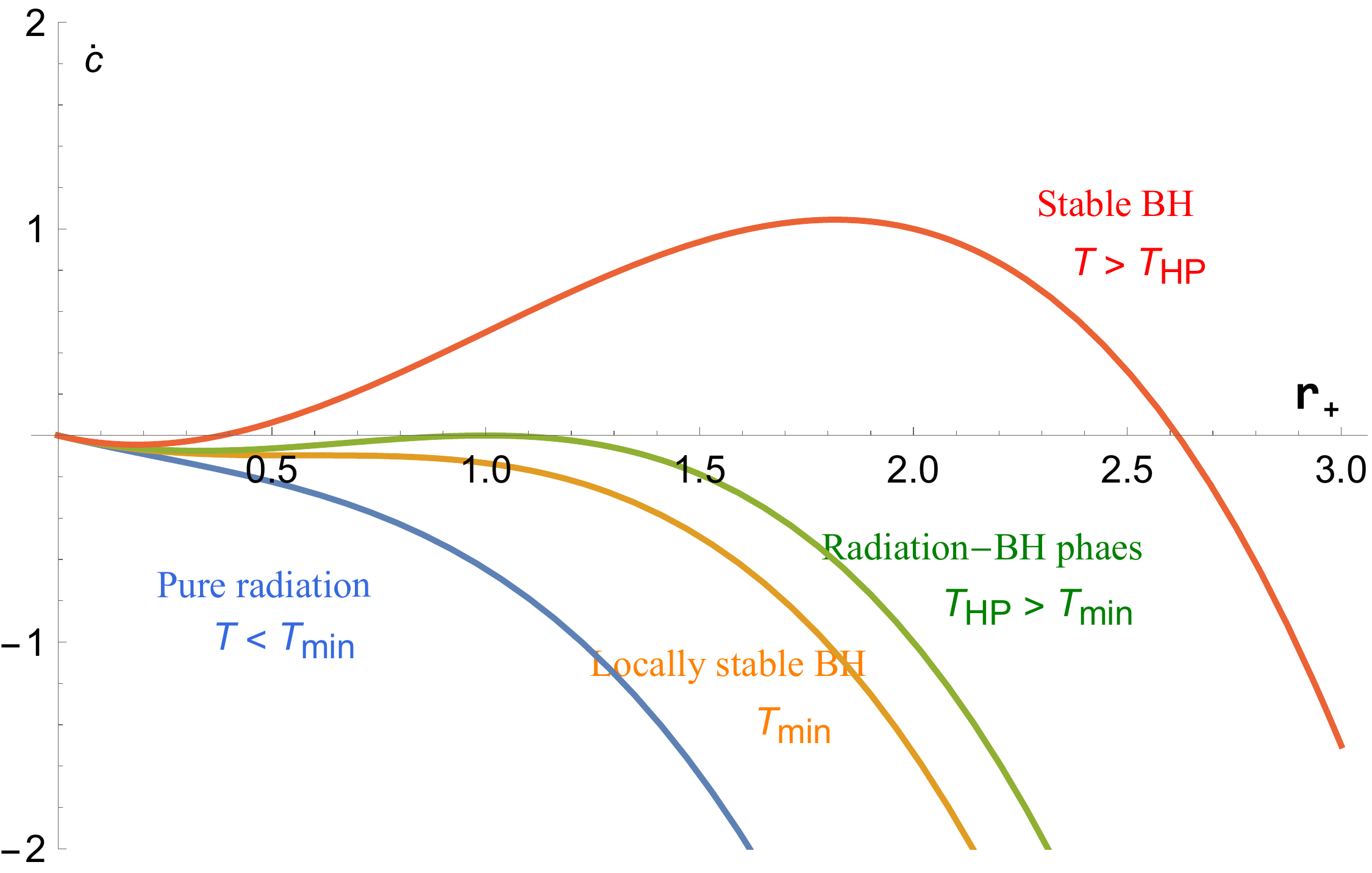}
    	\caption{\ $\dot{\mathcal{C}}$\ as a function of the black hole horizon radius for fixed l=1, and increasing temperature $ T=\frac{\sqrt{\frac{1}{2}}}{2\pi l} $ (blue line), $T=\frac{\sqrt{3}}{2\pi l}$ (orange line), $ T=\frac{1}{2\pi l} $ (green line) and $ T=\frac{3}{4\pi l} $ (red line) from top to bottom .}
    	\label{fig:car}
    \end{figure}
     where $M$ given by $M=\frac{r_{+}}{2}(1+\frac{r^{2}_{+}}{l^{2}})$ is the black hole mass and $\mathit{l}$ is the radius of curvature of the AdS space-time. The Hawking temperature at the horizon and the entropy is given by
     \begin{equation}
     T=\frac{{f}^{\prime }(r_{+})}{4\pi}=\frac{1}{4\pi r_{+}}\bigg(1+\frac{3r^{2}_{+}}{l^{2}}\bigg)\label{commutative8}\ ,~~~ \ S=\pi r_{+}^{2}.
     \end{equation}
     The form of $M$ shows that for any positive mass, there is only one horizon. As a consequence, this kind of black hole does not admit any extremal configuration in which $M$ has a minimum. From the formula (\ref{commutative8}) one can see that $T$ has a minimum value of $\frac{\sqrt{3}}{2\pi l}$ when $r_{+}=\frac{l}{\sqrt{3}}$. For $T<T_{min}$, there are no black holes but a pure radiation phase. The background heat bath is too cold to admit nucleation of black holes. For $T=T_{min}$, a single black hole is formed with a radius of $ r_{min}=\frac{l}{\sqrt{3}}$. For $T>T_{min}$, a pair of black holes (large/small) exist with radius given by
    \begin{equation}
    r_{l,s}=\frac{T}{2\pi T^{2}_{min}}\bigg(1\pm\sqrt{1-\frac{T^{2}_{min}}{T^{2}}}\bigg)\ ,~~~ \ r_{s}<r_{min} , \  r_{l}>r_{min} ,
    \end{equation}

    One can compute the difference of Euclidean action between the black hole metric and that of anti-de Sitter space, and in this case the contribution of the surface term is zero. The action equals to the difference in four-volumes of the two metrics and is given by\cite{Haking:1983}
    \begin{equation}\label{action}
      I=\frac{\pi r_{+}^{2}(l^{2}-r_{+}^{2})}{l^{2}+3r_{+}^{2}},
    \end{equation}
    Now we noticed
    \begin{equation}
     \dot{\mathcal{C}}\sim T \ln \mathcal{Z}=-TI_{E}=\frac{r^{3}_{+}}{4 l^{2}}-\frac{r_{+}}{4}\label{c1}.
    \end{equation}

   When the temperature is less than $T_{min}$, the maximum value of the complexity growth rate is at the origin $(r_{+}=0)$; When the temperature is equal to $T_{min}$, the function reach the inflection point at $r_{+} =\frac{1}{2 \pi T_{min}}=\frac{l}{\sqrt{3}}$. Above this temperature, there are two black holes, the small black hole corresponds to a locally minimum of $\dot{\mathcal{C}}$, while the larger one is locally stable being a locally maximum. With increasing T, $\dot{\mathcal{C}}$ becomes the locally maximum when the temperature reaches the Hawking phase transition temperature $T=\frac{1}{\pi l}\equiv T_{HP}$.

  \subsection{Reissner-Nordstrom AdS Black Holes}
      For many years, physicists have found that AdS charged black holes have almost the same thermodynamic properties as van der Waals gas (for example, have the same P-V criticality)\cite{Caldarelli:1999xj}. The standard 4-dimensional RN-AdS black hole metric is
     \begin{equation}
     ds^{2}=-\bigg(1-\frac{2M}{r}+\frac{r^{2}}{l^{2}}+\frac{Q^{2}}{r^{2}_{+}}\bigg)dt^{2}+\bigg(1-\frac{2M}{r}+\frac{r^{2}}{l^{2}}+\frac{Q^{2}}{r^{2}_{+}}\bigg)^{-1}dr^{2}+r^{2}d\Omega^{2}.
     \end{equation}
     The Hawking temperature at the horizon and the entropy is given by
     \begin{equation}
     T=\frac{{f}^{\prime }(r_{+})}{4\pi}=\frac{1}{4\pi r}\bigg(1+\frac{3r^{2}_{+}}{l^{2}}-\frac{Q^{2}}{r^{2}_{+}}\bigg)\ ,~~ \ S=\pi r_{+}^{2}.
     \end{equation}
     In order to obtain the partition function of the system, we calculate its Euclidean action. For a fixed charge $Q$, one considers a surface integral
     \begin{equation}\label{I_{s}}
     I_{s}=-\frac{1}{8\pi}\int_{\partial M}d^{3}x\sqrt{h}K-\frac{1}{4\pi}\int_{\partial M}d^{3}x\sqrt{h}n_{a}F^{ab}A_{b},
     \end{equation}
     The first term is the standard Gibbons-Hawking term and the second term is needed to impose fixed $Q$ as a boundary condition at infinity. So the total action is then given by
     \begin{equation}
     I=I_{EM}+I_{s}+I_{c},
     \end{equation}
     where $I_{EM}$ is given by $I_{EM}=-\frac{1}{16\pi}\int_{M}\sqrt{g}(R-F^{2}+\frac{6}{l^{2}})$, and $I_{c}$ represents the invariant counterterms needed to cure the infrared divergences \cite{Emparan:1999pm,Mann:1999pc}. The total action was first calculated in \cite{Chamblin:1999hg,Caldarelli:1999xj} and reads
     \begin{equation}\label{I}
     I=\frac{\beta}{4 l^2}\bigg(l^2r_{+}-r^3_{+}+\frac{3l^2Q^2}{r_{+}}\bigg) .
     \end{equation}
     \begin{figure}[htbp]
     	\centering
    	\includegraphics[width=0.7\linewidth, height=0.33\textheight]{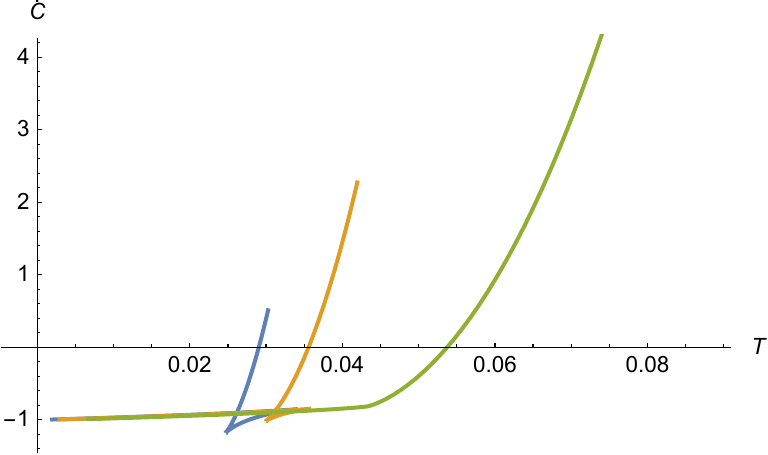}
    	\caption{\ $\dot{\mathcal{C}}$\ as a function of temperature for fixed Q =
  		1. The blue line corresponds to $P/P_{c} = 0.55$, and the green line corresponds to critical
  		pressure P = Pc $\approx $ 0.0033. Obviously, for $T < T_{c} \approx 0.043$
  		there is a (small black hole)-(large black hole) first-order phase transition.}
    	\label{fig:C-T}
     \end{figure}
     So in this case the complexity rate is
     \begin{equation}
     \dot{\mathcal{C}}\sim T \ln \mathcal{Z}=-TI_{E}=\frac{r^{3}_{+}}{4  l^{2}}-\frac{r_{+}}{4}-\frac{3Q^{2}}{4r_{+}}.
     \end{equation}
     Previous work on the critical behaviour of RN-AdS black hole in the non-extended phase space demonstrates that in the canonical (fixed charge) ensemble, for $Q<Q_{c}$, there exists a first order phase transition in the system \cite{Chamblin:1999hg,Chamblin:1999tk}. The critical point of the RN-AdS black hole, given by $T_{c}=\frac{\sqrt{6}}{18\pi Q}, P_{c}=\frac{1}{96\pi Q^{2}}$\cite{Kubiznak:2012wp}.

     We then consider the phase transition of the AdS charged black hole system in the extended phase space while we treat the black hole charge Q as a fixed external parameter, not a thermodynamic variable. The behaviour of $\dot{\mathcal{C}}$ is depicted in Fig.\ref{fig:C-T}. Since the $\dot{\mathcal{C}}$ demonstrates characteristic ``wallow tail'' behaviour, there is a first order transition in the system as $T<T_{c}$.

 \section{Relations to Lloyd Bound}
    \subsection{Neutral static black holes}
   Obviously, according to the definition of quantum complexity, we can see that any way to produce states has already limited the growth of complexity. Inspired by Margolus-Levitin theorem \cite{Margolus:1997ih}, Lloyd conjecture that the orthogonal time $\tau_{\perp}$ is bounded below by
    \begin{equation}
      \tau_{\perp}\geq \frac{h}{4E},
    \end{equation}
    where $E$ is the average energy of the state. If we take the reciprocal of both sides and describe the left side of the equation as the rate of complexity, then we came to the conclusion: the rate of complexity is limited by the energy of the system.
    \begin{equation}
    \dot{\mathcal{C}}\le\frac{2E}{\pi\hbar}.
    \end{equation}
    which is the Lloyd bound. In calculations with Schwarzschild-AdS black holes, $E$ will be the mass $M$ of the black hole. We found that
    \begin{figure}[htbp]
    	\centering
    	\includegraphics[width=0.7\linewidth, height=0.3\textheight]{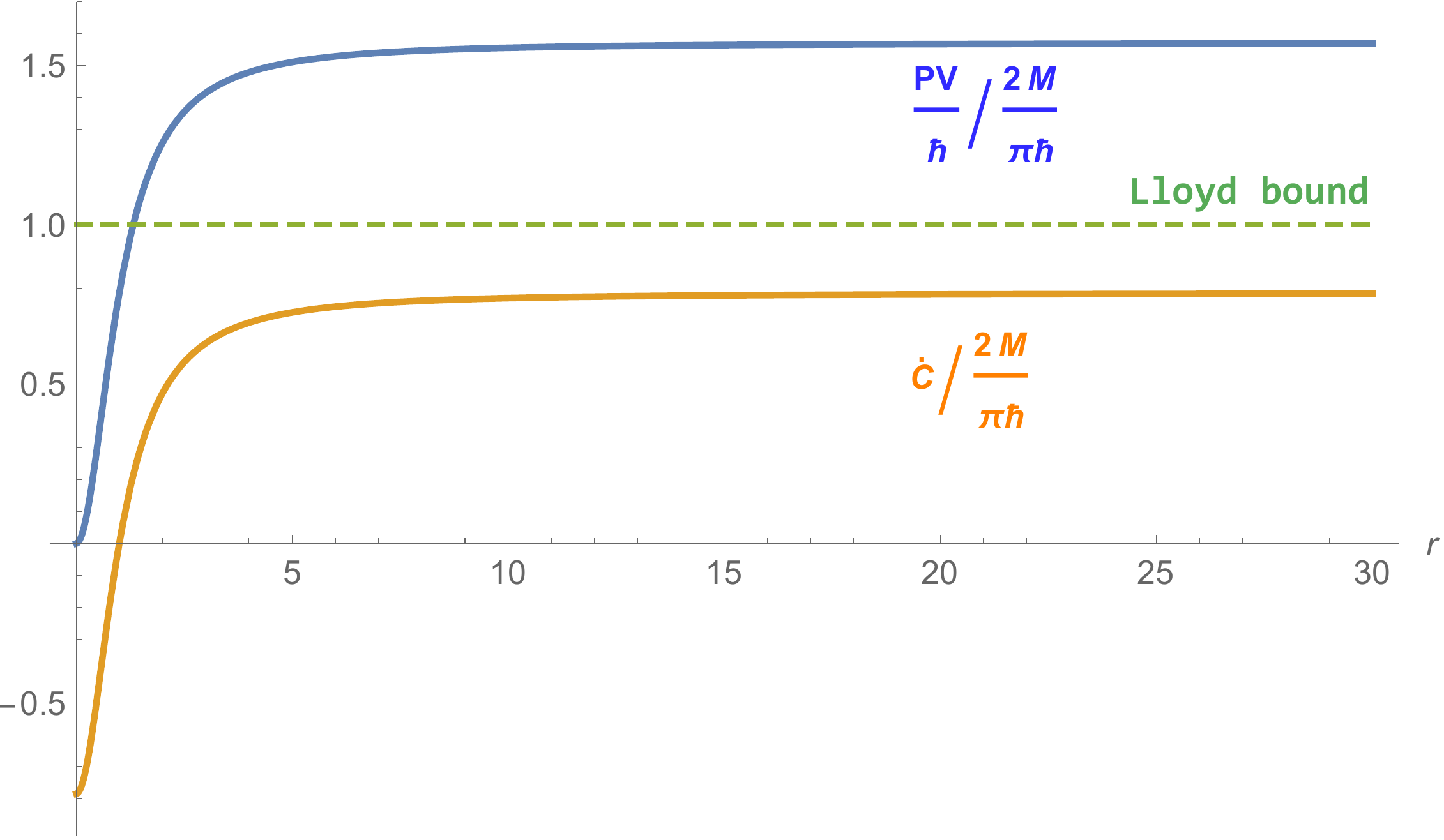}
    	\caption[Plot of $\frac{PV}{\hbar}/\frac{2M}{\pi \hbar}$ as a function of $r_{+}$]{Plot of $\frac{PV}{\hbar}/\frac{2M}{\pi \hbar}$ expressed with blue lines and $\dot{\mathcal{C}}/\frac{2M}{\pi \hbar}$ expressed with yellow lines as a function of the black hole radius $r_{+}$, we have set l=1, $\hbar=1$.}
    	\label{fig:ads-s}
    \end{figure}
     CV2.0 does not strictly obey the Lloyd bound. In Fig.3, we plot $\frac{PV}{\hbar}/\frac{2M}{\pi \hbar}$ and $\dot{\mathcal{C}}/\frac{2M}{\pi \hbar}$ as a function of the black hole radius $r_{+}$.  We can see that in this case the complexity rate denoted by (\ref{c1}) always satisfies the Lloyd bound. Actually, we test the Schwarzschild AdS black hole of all sizes $r_{+}\sim l_{ads}$, $r_{+} \ll l_{ads}$, $r_{+} \gg l_{ads}$. They are always consistent with the Lloyd bound.

    \subsection{Charged black holes}
    As argued in \cite{Brown:2015lvg}, the existence of conserved charges slows down the growth of complexity at late time,
    The thermofield double state includes a chemical potential $\mu$:
    \begin{equation}
    \left| TFD_{\mu}\right\rangle=\frac{1}{\sqrt{Z}}\sum _{n}e^{-\beta(E_{n}+\mu Q_{n})/2} \left|E_{n}Q_{n}\right\rangle _{L}\left|E_{n}-Q_{n}\right\rangle _{R}.
    \end{equation}
    This state time-evolves by the Hamiltonian $H_{L}+\mu Q_{L}$ on the left, and $H_{R}-\mu Q_{R}$ on the right:
    \begin{equation}
    \left|\psi(t_{L},t_{R})\right\rangle=e^{-i(H_{L}+\mu Q_{L})t_{L}}e^{-i(H_{R}-\mu Q_{R})t_{R}}\left|TFD_{\mu}\right\rangle ,
    \end{equation}
     where
     $H_{L}$ and $H_{R}$ are the $\mu = 0$ Hamiltonians. According to the same argument leading to the boundary of $\mu = 0$, the complex boundary becomes \cite{Brown:2015lvg}
    \begin{equation}
    \dot{\mathcal{C}}\le\frac{2}{\pi\hbar}\left[(M-\mu Q)-(M-\mu Q)_{gs}\right]\label{commutative11} ,
    \end{equation}
    where $(M-\mu Q)_{gs}$ is the ground state of $(M-\mu Q)$, which is either an empty AdS spacetime or an extreme black hole. In Fig.6, we plot the growth of the complexity as a function of the black hole horizon for the three radius. All of them  are below the bound given in (\ref{commutative11}).
\begin{figure}[htbp]
    	\centering
    	\subfigure{}
        \begin{minipage}{4.7cm}
        \centering
    	\includegraphics[scale=0.26]{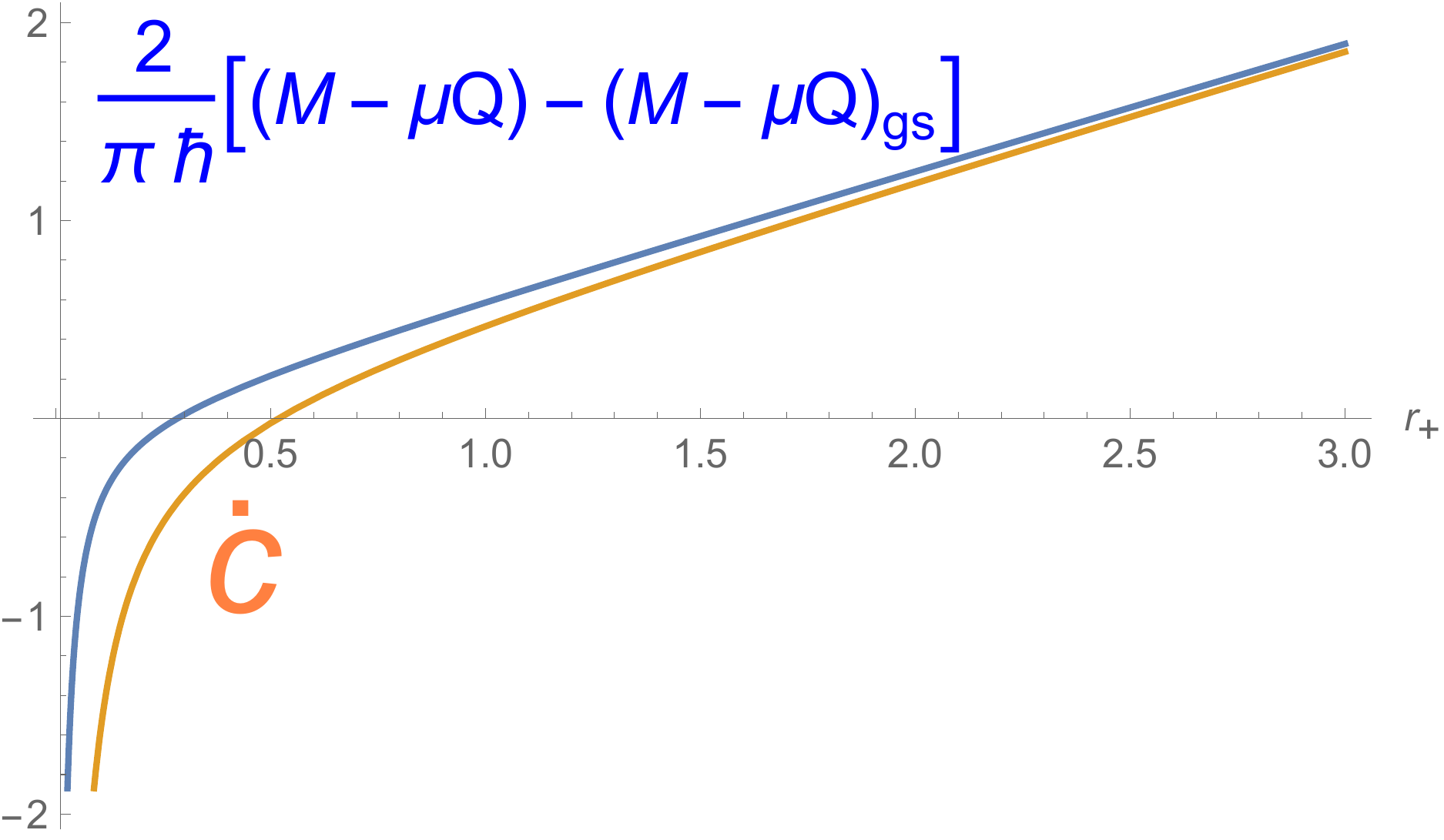}
        \end{minipage}
        \subfigure{}
        \begin{minipage}{4.7cm}
     	\centering
    	\includegraphics[scale=0.26]{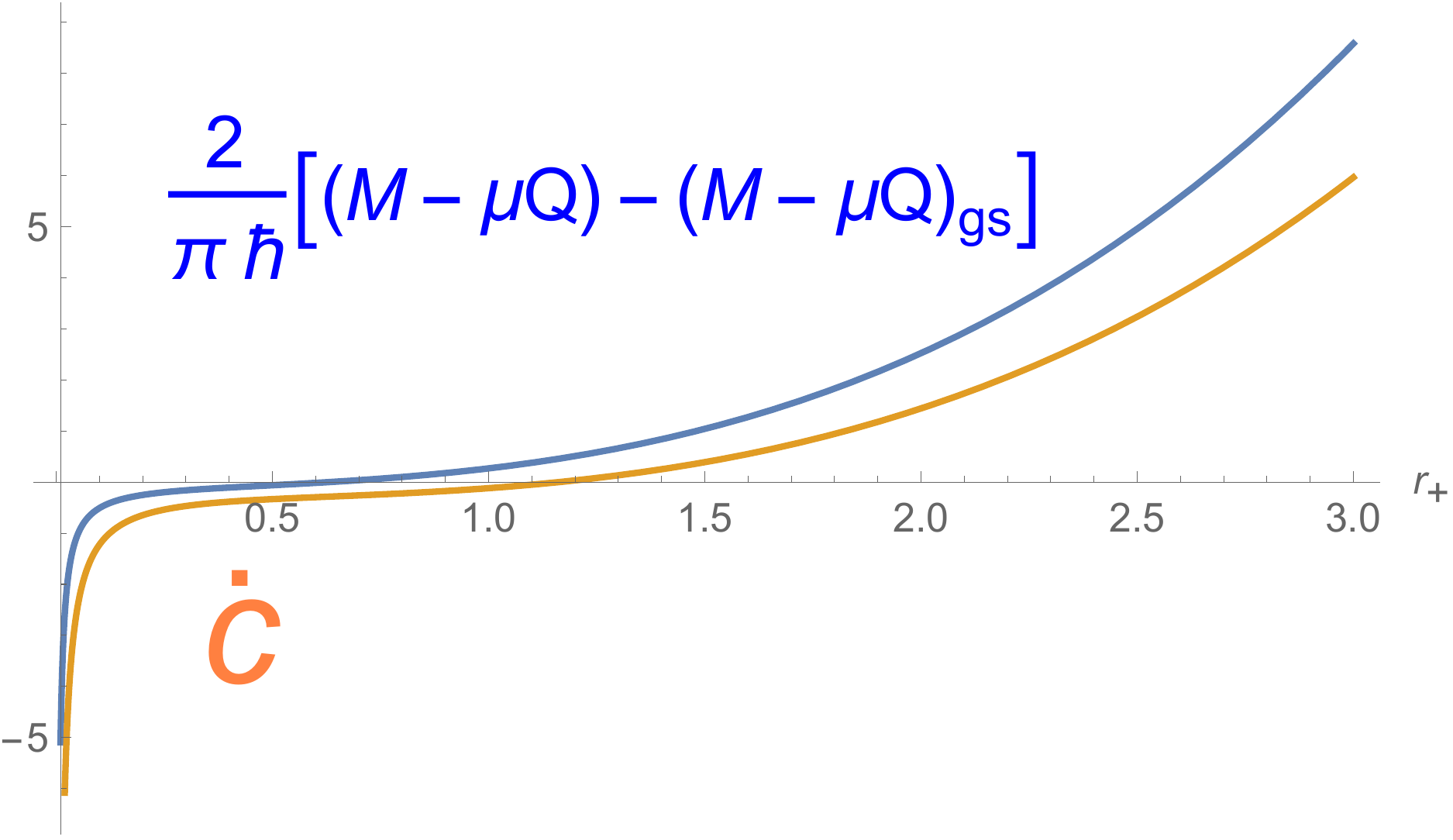}
        \end{minipage}
        \subfigure{}
        \begin{minipage}{4.7cm}
       	\centering
        \includegraphics[scale=0.26]{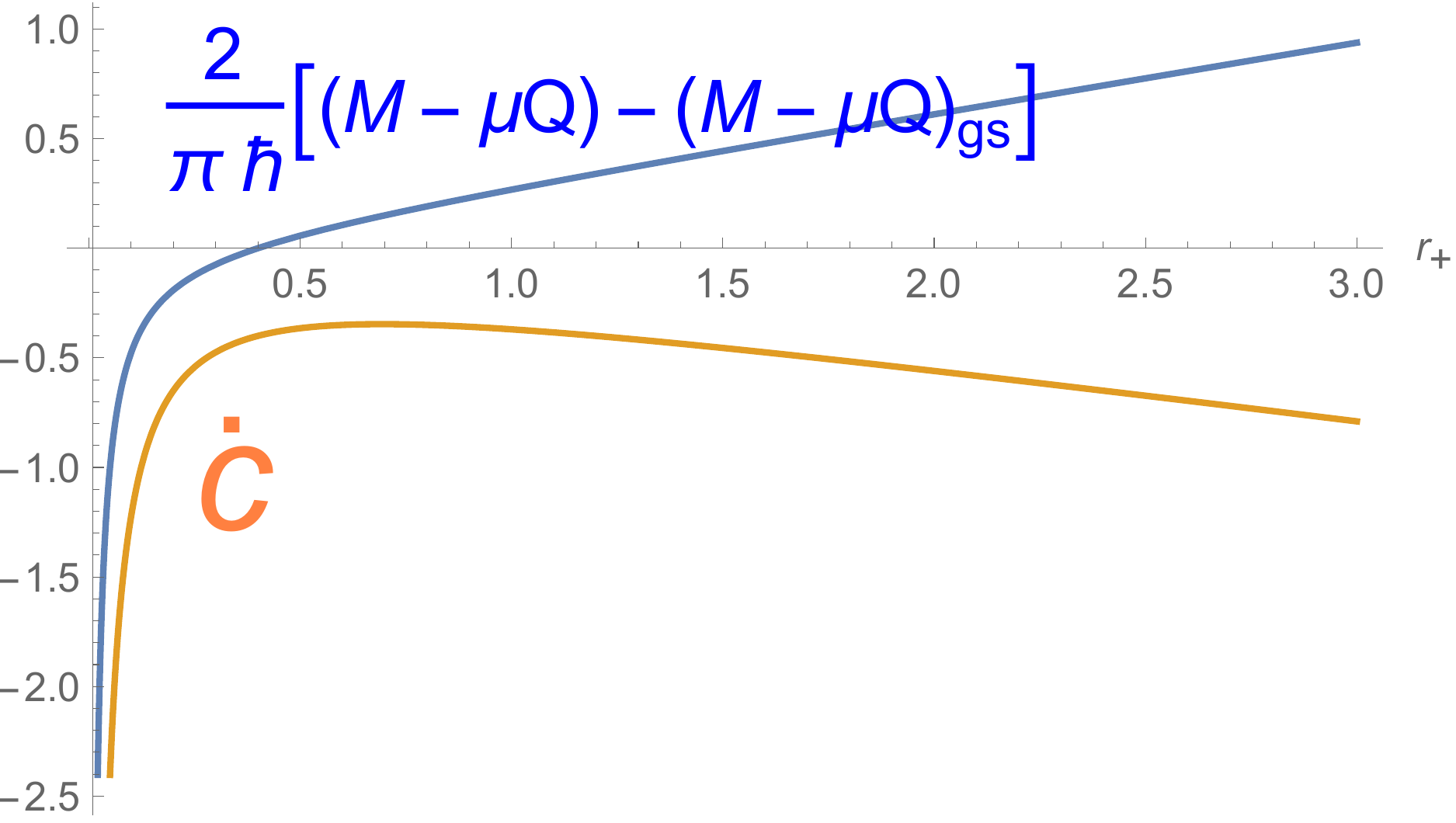}
        \end{minipage}
    	\caption{$\dot{\mathcal{C}}$ expressed with yellow lines and Lloyd bound  expressed with blue lines as a function of $r_{+}$ with different sizes. The left figure is the case of the $r_{+}\sim l_{AdS}$, the middle figure is the case of the $r_{+} >> l_{AdS}$, and the figure on the right is the case of the $r_{+} << l_{AdS}$. $l=1,Q=0.4$}
    	\label{01}
    \end{figure}

    \subsection{Einstein scalar theory: case 1 }
    In \cite{Liu:2019mxz}, the authors  found that for black holes in Einstein-scalar theory the Lloyd bound can be violated, because the volume of black hole singularity becomes negative. It is of interesting to check whether such black holes still violates the Lloyd bound in own set-up. The Lagrangian of the Einstein-scalar theory takes the general form
\begin{equation}\label{L}
  \mathcal{L}=\sqrt{-g}\left(R-\frac{1}{2}(\partial\phi)^{2}-V(\phi)\right).
\end{equation}
We begin with the example D = 4 case, in which  the potential is \cite{Zloshchastiev:2004ny}
\begin{equation}\label{V1}
  V(\phi)=-2g^{2}\left((\cosh\phi+2)-2\beta^{2}(2\phi+\phi\cosh\phi-3\sinh\phi)\right),
\end{equation}
where the parameter $\beta$ is a fixed dimensionless quantity. The theory admits asymptotic AdS black hole, given by \cite{Zloshchastiev:2004ny}
\begin{equation*}
  ds^{2}=-fdt^{2}+\frac{dr^{2}}{f}+r(r+q)d\Omega^{2}_{2,k},\ ~~~ \  e^{\phi}=1+\frac{q}{r},
\end{equation*}
\begin{equation}
  f=g^{2}r^{2}+k-\frac{1}{2}g^{2}\beta^{2}q^{2}+g^{2}(1-\beta^{2})qr+g^{2}\beta^{2}r^{2}(1+\frac{q}{r})\log(1+\frac{q}{r}).
\end{equation}
The solution contains only one integral constant $q$, parameterizing the mass
\begin{equation}
  M=\frac{1}{12}g^{2}\beta^{2}q^{3},
\end{equation}
and the thermodynamical variables are \cite{Liu:2019mxz}
\begin{equation}
 T=\frac{f^{\prime}(r_{+})}{4\pi},\ ~~~ \ S=\pi r_{+}(r_{+}+q), \ ~~~ \ P=\frac{3g^{2}}{8\pi},
\end{equation}
\begin{equation*}
V=\frac{2}{3}\pi r_{+}^{3}(1+\frac{q}{r_{+}})(2+\frac{q}{r_{+}})\left(1+\beta^{2}\log(1+\frac{q}{r_{+}})\right)-\frac{1}{9}\pi\beta^{2}q(q^{2}+12qr_{+}+12r_{+}^{2}),
\end{equation*}
combined with the above physical quantities, we calculate the complexity as
\begin{equation}
 \dot{\mathcal{C}}=\frac{1}{12}g^{2}\left(-\beta^{2}q^{3}+3r_{+}(q+r_{+})      \left(q-2\beta^{2}q+2r_{+}+\beta^{2}(q+2r_{+})\log[\frac{q+r_{+}}{r_{+}}] \right)\right).
\end{equation}
Note that since $T>0$, $S>0$ and $M>0$, we can conclude that $\dot{\mathcal{C}}=TS-M<2M$, so the Lloyd bound is satisfied for this case.
\subsection{Einstein scalar theory: case 2 }
We now consider the other scalar potential as given in \cite{feng13}
\begin{equation*}
\begin{split}
 V(\phi)=&-\frac{1}{2}(D-2)g^{2}e^{\frac{\mu-1}{\nu} \Phi }\\
 &\times\left[(\mu-1)((D-2)\mu-1)e^{\frac{2}{\nu}\Phi}-2(D-2)(\mu^{2}-1)e^{\frac{1}{\nu}\Phi}+(\mu+1)
 ((D-2)\mu+1)\right] \\
 &-\frac{(D-3)^{2}}{2(3D-7)}(\mu+1)\alpha e^{-\frac{1}{\nu}(4+\frac{\mu+1}{D-3})\Phi}(e^{\frac{1}{\nu}\Phi}-1)^{3+\frac{2}{D-3}}\times [(3D-7)e^{\frac{1}{\nu}\Phi} \\
 & _{2}F_{1} [2,1+\frac{(D-2)(\mu+1)}{D-3};3+\frac{2}{D-2};1-e^{\frac{1}{\nu}\Phi}]  \\
 & -((3D-7)+(D-2)(\mu-1)) _{2}F_{1} [3,2+\frac{(D-2)(\mu+1)}{D-3};4+\frac{2}{D-2};1-e^{\frac{1}{\nu}\Phi}]
\end{split}
\end{equation*}
The metric of the solution is given by \cite{feng13}
\begin{equation}
  ds^{2}=-\frac{f}{H^{1+\mu}}dt^{2}+H^{\frac{1+\mu}{D-3}}\left(\frac{dr^{2}}{f}+r^{2}d\Omega^{2}_{D-2}\right),\ ~~~ \ H=1+\frac{q}{r^{D-3}}
\end{equation}
\begin{equation*}
\begin{split}
 f=&g^{2}r^{2}H^{\frac{(D-2)(\mu+1)}{D-3}}+kH-\beta g^{2}r^{2}(H-1)^{\frac{D-1}{D-3}}\\
 &\times_{2}F_{1}(1,\frac{(D-2)(\mu+1)}{D-3};\frac{2(D-2)}{D-3};1-\frac{1}{H}).
\end{split}
\end{equation*}
The solution contains one integral constant q, parameterizing the mass of the solution, given by
\begin{equation}
  M=\frac{(D-2)\Omega_{D-2}q(\beta g^{2}q\frac{2}{D-3}+k\mu)}{16\pi},
\end{equation}
thermodynamical variables are given by \cite{Liu:2019mxz}
\begin{equation*}
T=\frac{H_{+}^{-\frac{(D-2)(\mu+1)}{2(D-3)}}}{4\pi}f^{\prime}_{+},\ ~~~ \ S=\frac{\Omega_{D-2}}{4}r_{+}^{D-2}H_{+}^{\frac{(D-2)(\mu+1)}{2(D-3)}}\ ~~~ \ p=\frac{(D-1)(D-2)g^{2}}{16\pi},
\end{equation*}
\begin{equation}
\begin{split}
&V=\frac{\Omega_{D-2}r^{D-1}((1-\mu)H+\mu+1)H^{\frac{(D-2)(\mu+1)}{2(D-3)}}}{2(D-1)}+\frac{\Omega_{D-2}\beta q^{\frac{D-1}{D-3}}}{2(D-1)H} \\
&\left(2H-((1-\mu)(H-1)+2)_{2}F_{1}[1,\frac{(1+\mu)(D-2)}{D-3}; \frac{2(D-2)}{D-3};1-H^{-1}]\right).
\end{split}
\end{equation}
Simply, we set $D=4, \beta=g=k=1$. In the case of $\mu=-1$, we calculate the complexity result as
\begin{equation}
 \dot{\mathcal{C}}= \frac{(2r_{+}^{3}-q^{3}+q)\Omega_{2}}{16\pi},
\end{equation}
and for $\mu=1$, the form is expressed as
\begin{equation}
 \dot{\mathcal{C}}=\frac{(2r_{+}^{3}-5q^{3}+q(4r_{+}^{2}-3))\Omega_{2}}{16\pi}.
\end{equation}
Note also that in case, we can also conclude that the Lloyd bound is satisfied because  $\dot{\mathcal{C}}=TS-M<2M$.

 From these results, we may conclude that the complexity/partition function relation can satisfy the Lloyd bound for various conditions. Under the consideration of the principle of maximal work, the Lloyd bound seems not to be saturated by the complexity/partition function relation. For example, for Schwarzschild-AdS black holes with planar horizons the complexity growth rate $\dot{\mathcal{C}}=M/2$, having a factor difference from the Lloyd bound.

 \section{ Discussion and Conclusion}

      In summary, by examining the CV 2.0 (i.e. $\mathcal{\dot{C}}\sim PV$) and using a fundamental relation between thermodynamics and statistical physics $ \ln \mathcal{Z}=PV/kT=-\Omega/kT$, we obtain a relation between the complexity, the grand potential and the partition function. For canonical ensembles, the grand potential reduces to the free energy of the system.
      In order to illustrate the validity of our proposal $\mathcal{\dot{C}}\sim T \ln \mathcal{Z}$, we have studied the complexity of the TFD state of various deformations of the SYK model. For the original SYK model and its extension to the complex field, the relation $\mathcal{\dot{C}}\sim T S$ is well respected. For $(2+1)$-dimensional SYK model with two bands where $f$ fermions form the bulk geometry while the conducting $c$ fermions live on the boundary. It turns out that only the $f$-fermions involving the SYK interactions obeys the relation $\mathcal{\dot{C}}\sim T S$.
      We further studied the complexity growth rate of the TFD state of the SYK ``wormholes" and found that the result agrees with \cite{JTcomplexity} at the qualitative level.

      We then applied the complexity/partition function relation to cases of AdS black holes. For both Schwarzschild-AdS and RN-AdS black holes, the relation $\mathcal{\dot{C}}\sim -F$ holds. The connections between complexity and phase transition were then discussed. It seems that the quantum complexity can be regarded as an order parameter for phase transitions.
      We have also checked whether our proposal violates the Lloyd bound. The results show that both the original Lloyd bound and the generalized Lloyd bound are satisfied for Schwarzschild-AdS and RN-AdS black holes.

       Quantum computational process can be considered as a way toward complexity increasing and work should be done during this process. In thermodynamics, there is a well-known principle-the principle of  maximum work $dW\leq -dF$.
      In \cite{huang2017}, it was proposed that for general two-horizon black holes, the complexity growth rate in the WDW patch can be expressed as $\mathcal{\dot{C}}={H}_{+}-{H}_{-} $, that is to say, the difference between the enthalpy associated with the inner and outer horizons. Later, a new CV conjecture was proposed as \cite{Liu:2019mxz}
      \be
      \mathcal{\dot{C}}=2P \Delta V,
      \ee
      with $\Delta V=V^{+}-V^{-}$, where $V^{\pm}$ are the thermodynamical volumes on the outer and inner horizons. This is analogous to the  mechanical work in the ordinary thermodynamics, i.e. $\Delta W=P \Delta V$. From the principle of  maximum work, we can write down an analogous relation $\mathcal{\dot{C}}= 2 P \Delta V=2 \Delta W\leq -2 \Delta F$. For CV 2.0, this relation simply reduces to
      $\mathcal{\dot{C}}\leq -\Delta F$. In this work, we assume $\mathcal{\dot{C}}\sim -\Omega$ and examine this relation for several conditions. The results show that the Lloyd bound is not violated, but cannot be saturated exactly. Much more can be studied on the relation between complexity, thermodynamics and statistical mechanics, we defer further study on their connections in future studies.

 \section*{Acknowledgement}
  We would like to thank Hong L$\rm \ddot{u}$, Song He and Runqiu Yang for helpful discussions. This work is partly supported by NSFC (No.11875184 $\&$ No.11805117).

\end{document}